\documentclass[prd,twocolumn,superscriptaddress,nofootinbib]{revtex4-2}
\usepackage{amsmath,amssymb,graphicx,booktabs}

\begin{document}

\title{The Goldstone--Wilczek Charge on the Rim}

\author{Laurence A. Jacobs}
\affiliation{University of Zurich, Zurich, Switzerland}
\affiliation{Center for Complexity Sciences, National University of Mexico, Mexico City, Mexico}

\date{October 31, 2026 \\[8pt] \url{https://doi.org/10.5281/zenodo.23974269}}

\begin{abstract}
A solenoid flux $\Phi=2\pi\alpha$ threaded through the center of the Kaplan--Sen lattice disk binds a ground-state charge $Q_{\rm core}=\alpha-\mathrm{round}(\alpha)$, and the edge carries exactly the opposite charge. The edge charge is the Goldstone--Wilczek charge of the edge ladder, $Q_{\rm edge}=-\tfrac12\eta$, with $\eta$ its spectral asymmetry. At $\Phi=\pi$ one unit of charge crosses the disk. We ask how the rim operator, the Schur complement of the disk Hamiltonian on its boundary sites, registers this transfer. Its count of negative eigenvalues never changes over the flux cycle. Instead, the unit of charge enters the rim through a simple pole: the extremal eigenvalue of the rim operator passes through infinity in the same flux interval as the charge jump, with $\Delta Q_{\rm edge}=\tfrac12\Delta\,\mathrm{sign}\,\lambda_{\rm ext}$. The residue factorizes into the coupling of the inner edge mode to the rim and the inverse slope of that mode's level, and grows linearly with the radius. All statements are checked numerically on disks of radius $R=11$, $19$ and $34$.
\end{abstract}

\maketitle

\noindent\emph{In memory of Jeffrey Goldstone (1933--2026).}

\section{Introduction}

Goldstone and Wilczek showed that a fermion coupled to a slowly varying background acquires an induced current, and with it a charge, fixed by the topology of the background rather than by its details~\cite{GoldstoneWilczek1981}. The fractional charge of the Jackiw--Rebbi soliton~\cite{JackiwRebbi1976} is the simplest case; the spectral asymmetry of Atiyah, Patodi and Singer~\cite{APS1975} is the general bookkeeping. On a domain wall the same induced current is the inflow of Callan and Harvey~\cite{CallanHarvey1985}, which carries charge between the bulk and the boundary, and Laughlin's flux-threading argument~\cite{Laughlin1981} is its most familiar form.

Kaplan and Sen recently proposed a lattice disk whose boundary carries a single chiral mode~\cite{KaplanSen2024}. Golterman and Shamir analyzed such constructions through an effective boundary operator~\cite{GoltermanShamir2026}, and in Ref.~\cite{JacobsRim2026} we defined the rim operator of the Kaplan--Sen disk, the Schur complement of the Hamiltonian on the boundary sites, and showed that it has a pole at flux $\Phi=\pi$. Aoki and Fukaya had found that at $\Phi=\pi$ a zero mode localized at the flux coexists with the edge zero mode~\cite{AokiFukaya2023}.

This note connects these observations. We thread a flux through the disk, follow the ground-state charge, and identify it with the Goldstone--Wilczek charge of the edge ladder. We then show that the rim operator, whose spectrum is blind to the transfer of charge at $\Phi=\pi$, records it through its pole.

\section{Setup}

We use the Wilson Hamiltonian of Ref.~\cite{KaplanSen2024},
\begin{equation}
H=\gamma_0 D,\qquad D=\sum_{\mu=1,2}\gamma_\mu\partial_\mu+M+\frac r2\Delta,
\end{equation}
with $\gamma_0=\sigma_3$, $\gamma_{1,2}=\sigma_{1,2}$, symmetric lattice derivatives, $M=r=1$, and open boundary conditions on the sites of a square lattice inside a circle of radius $R$. A flux $\Phi=2\pi\alpha$ through the central plaquette enters in string gauge, so that $H(\alpha+1)=H(\alpha)$ identically.

The rim is the set of sites with a nearest neighbor outside the disk; the bulk is the rest. With $H_{rr}$, $H_{rb}$, $H_{bb}$ the corresponding blocks, the rim operator is~\cite{JacobsRim2026}
\begin{equation}
H_{\rm eff,rim}=H_{rr}-H_{rb}H_{bb}^{-1}H_{br}=\left(P_{\rm rim}H^{-1}P_{\rm rim}\right)^{-1},
\label{eq:rim}
\end{equation}
which is the free-field form of the Golterman--Shamir operator restricted to rim fields~\cite{GoltermanShamir2026}.

We fill the Dirac sea, all states with $w<0$, and measure the charge relative to $\alpha=0$ in three regions: the core, $r<R/2$; the edge band, $r>R-4$; and the rest.

\section{The Goldstone--Wilczek charge}

In the continuum the edge modes of a disk with flux $\alpha$ through its center form the ladder~\cite{AokiFukaya2023}
\begin{equation}
w_n=\frac vR\left(n+\tfrac12-\alpha\right),\qquad n\in\mathbb Z,
\label{eq:ladder}
\end{equation}
the half-integer shift coming from the induced spin connection. Its spectral asymmetry, regularized with the Hurwitz zeta function, is
\begin{equation}
\eta(\alpha)=\lim_{s\to0}\sum_n\mathrm{sign}(w_n)\,|w_n|^{-s}=2\left(\alpha-\mathrm{round}(\alpha)\right),
\end{equation}
and the charge of the filled ladder relative to $\alpha=0$ is the Goldstone--Wilczek charge
\begin{equation}
Q_{\rm edge}=-\tfrac12\,\eta(\alpha)=-\left(\alpha-\mathrm{round}(\alpha)\right).
\label{eq:GW}
\end{equation}
Here $\mathrm{round}(\alpha)$ is the integer nearest to $\alpha$, so that $\alpha-\mathrm{round}(\alpha)$ is a sawtooth in $(-\tfrac12,\tfrac12)$ that jumps by $-1$ at half-integer $\alpha$.
Charge conservation then requires the core to bind $Q_{\rm core}=\alpha-\mathrm{round}(\alpha)$: the fractional part of the Hall response $\Phi/2\pi$, with the integer fixed by which side of $\Phi=\pi$ one is on. At $\Phi=\pi$ the sawtooth jumps, and one unit of charge crosses between core and edge through the pair of zero modes found in Ref.~\cite{AokiFukaya2023}.

\begin{figure*}
\includegraphics[width=0.9\textwidth]{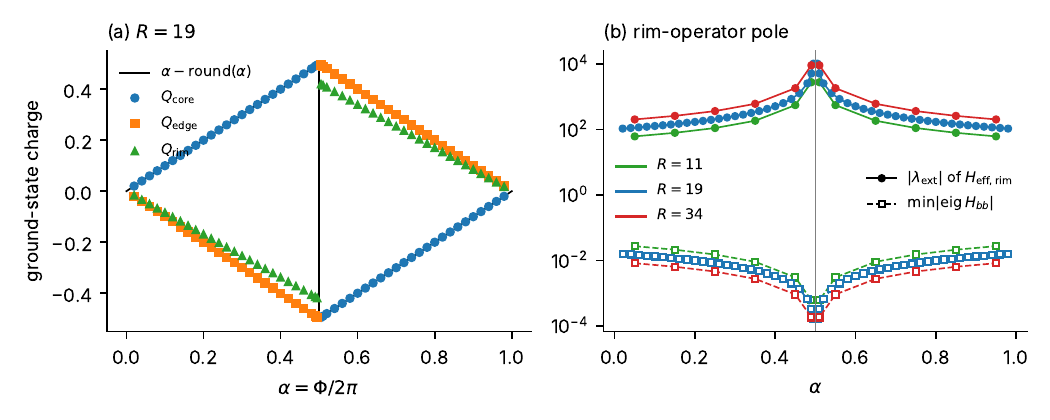}
\caption{(a) Ground-state charge in the core, the edge band and the outermost layer of sites ($Q_{\rm rim}$) of the $R=19$ disk as a function of the flux, with the Goldstone--Wilczek sawtooth, Eq.~\eqref{eq:GW}. (b) The pole of the rim operator for $R=11$, $19$ and $34$: the extremal eigenvalue $|\lambda_{\rm ext}|$ of $H_{\rm eff,rim}$ diverges at $\alpha=\tfrac12$ as the lowest level of the bulk block, $\min|{\rm eig}\,H_{bb}|$, goes to zero; their product is constant along the flux cycle (Table~\ref{tab:pole}).}
\label{fig:main}
\end{figure*}

Table~\ref{tab:charge} and Fig.~\ref{fig:main}(a) show that the lattice disk obeys Eq.~\eqref{eq:GW} to high accuracy. The core charge agrees with the sawtooth to $3\times10^{-12}$ on the largest disk, the edge band carries the opposite charge to better than $5\times10^{-5}$, and the total charge stays at zero to $10^{-13}$. The outermost layer of sites carries a fixed fraction of the edge charge, $0.85$ at all three radii, which is the weight of the edge mode on that layer.

\begin{table}
\caption{Ground-state charge bookkeeping: $\delta_{\rm core}=\max|Q_{\rm core}-(\alpha-\mathrm{round}(\alpha))|$ and $\delta_{\rm edge}=\max|Q_{\rm edge}+Q_{\rm core}|$. Maxima are over the flux grid ($52$ points for $R=19$, $12$ otherwise), which avoids $\alpha=\tfrac12$.}
\label{tab:charge}
\begin{ruledtabular}
\begin{tabular}{cccc}
$R$ & $\delta_{\rm core}$ & $\delta_{\rm edge}$ & $\max|Q_{\rm tot}|$\\
\hline
11 & $1.0\times10^{-4}$ & $7.8\times10^{-5}$ & $2.5\times10^{-14}$\\
19 & $2.9\times10^{-7}$ & $4.9\times10^{-5}$ & $2.2\times10^{-13}$\\
34 & $2.9\times10^{-12}$ & $3.4\times10^{-5}$ & $4.1\times10^{-13}$\\
\end{tabular}
\end{ruledtabular}
\end{table}

\section{The charge enters through the pole}

The rim operator sees none of this in its spectral count. Over the whole flux cycle its number of negative eigenvalues stays at $N_{\rm neg}=N_{\rm rim}$ ($60$, $104$ and $192$ for the three radii), even as a unit of charge moves onto the edge. By Haynsworth's inertia additivity~\cite{Haynsworth1968},
\begin{equation}
N_{\rm neg}(H)=N_{\rm neg}(H_{bb})+N_{\rm neg}(H_{\rm eff,rim}),
\end{equation}
and since the ground-state charge $Q_{\rm tot}=\Delta N_{\rm neg}(H)$ is also constant, the bulk count is constant as well. The bulk, which is a smaller disk with its own edge and core, must therefore cross zero at $\Phi=\pi$ with a compensating pair of levels, just as the full disk does.

The transfer is recorded instead by the inverse in Eq.~\eqref{eq:rim}. Writing $H_{bb}^{-1}$ in its spectral form, the level $\varepsilon_0(\alpha)$ of the bulk inner edge that crosses zero at $\alpha=\tfrac12$ gives
\begin{equation}
H_{\rm eff,rim}\simeq-\frac{H_{rb}|\psi_0\rangle\langle\psi_0|H_{br}}{\varepsilon_0(\alpha)}+\text{regular},
\end{equation}
so that the extremal eigenvalue is
\begin{equation}
\lambda_{\rm ext}\simeq\frac{g^2}{s\,(\alpha-\tfrac12)},\qquad g^2=\|H_{rb}\psi_0\|^2,\quad \varepsilon_0\simeq-s\left(\alpha-\tfrac12\right).
\label{eq:pole}
\end{equation}
As $\alpha$ passes through $\tfrac12$ this eigenvalue passes through infinity and changes sign while all other eigenvalues stay finite.

The numerics confirm Eq.~\eqref{eq:pole} and tie it to the charge. In each disk the residue $c=\lambda_{\rm ext}(\alpha-\tfrac12)$ is constant over the flux cycle to a relative spread below $6\times10^{-4}$, and the sign of $\lambda_{\rm ext}$ flips in the same flux interval as the jump of $Q_{\rm edge}$, with
\begin{equation}
\Delta Q_{\rm edge}=\tfrac12\,\Delta\,\mathrm{sign}\,\lambda_{\rm ext}=+1 .
\end{equation}
Table~\ref{tab:pole} shows the factorization of the residue. The slope of the bulk level scales as $s=v/(R-1)$ with $v\simeq0.60$ fixed to $1\%$, the radius of the bulk disk being $R-1$. The coupling $g^2\simeq1.68$, measured as $|\lambda_{\rm ext}|\min|{\rm eig}\,H_{bb}|$, is constant to $2\%$. Its size, of order one, is what the inner edge of the bulk disk gives, one lattice spacing from the rim; the core-localized partner of the pair, far from the rim, would couple to it only exponentially weakly. The residue therefore grows as
\begin{equation}
c=\frac{g^2}{v}\,(R-1),
\end{equation}
to within $3\%$ over the three radii (Table~\ref{tab:pole}): the pole sharpens with the radius because the inner edge level flattens.

\begin{table}
\caption{Factorization of the pole residue $c=\lambda_{\rm ext}(\alpha-\tfrac12)$. $s$ is the slope of $\min|{\rm eig}\,H_{bb}|$ in $|\alpha-\tfrac12|$.}
\label{tab:pole}
\begin{ruledtabular}
\begin{tabular}{ccccc}
$R$ & $c$ & $c/(R-1)$ & $s\,(R-1)$ & $g^2$\\
\hline
11 & 27.61 & 2.76 & 0.607 & 1.677\\
19 & 50.93 & 2.83 & 0.599 & 1.696\\
34 & 90.83 & 2.75 & 0.605 & 1.665\\
\end{tabular}
\end{ruledtabular}
\end{table}

\section{Discussion}

Two facts here are not new. Charge bound to flux and its transfer to the edge are the content of Laughlin's argument and of anomaly inflow~\cite{Laughlin1981,CallanHarvey1985}, and the pair of zero modes at $\Phi=\pi$ was found by Aoki and Fukaya~\cite{AokiFukaya2023}. What this note adds is the reading of the edge charge as a Goldstone--Wilczek charge of the ladder~\eqref{eq:ladder}, with the sawtooth reproduced by the lattice to $3\times10^{-12}$, and the identification of the rim operator's pole as the place where that charge enters the boundary description.

The second point has a simple moral. A boundary operator obtained by eliminating the bulk, as in Eq.~\eqref{eq:rim}, is exact for the boundary propagator, yet its spectrum cannot follow a unit of charge onto the boundary: by inertia additivity its negative count is fixed. The charge comes in through infinity. Any proposal to define chiral boundary dynamics from such an operator alone~\cite{KaplanSen2024,GoltermanShamir2026} must account for this pole, whose residue grows with the size of the system.

Everything here is free, two-dimensional and quadratic. Whether the gauge field, integrated over its boundary values, changes the picture is the open question raised in Refs.~\cite{KaplanSen2024,GoltermanShamir2026} and is not addressed.

\section*{A personal note}

Jeffrey Goldstone died on 12 October 2026. With Sidney Coleman, he was the most elegant physicist I have ever known. One episode shows the kind of elegance I mean.

In 1975 I took the Kerman--Klein method that Jeffrey and Roman Jackiw had developed for quantized nonlinear waves~\cite{GoldstoneJackiw1975} to the next order in the coupling, computing the mass of the $\lambda\phi^4$ soliton in three different ways~\cite{Jacobs1976}. The Kerman--Klein/Goldstone--Jackiw route was exceedingly messy. To keep track of the terms, several hundred as I recall, I devised a graphical bookkeeping. The calculation took me months, and I made many mistakes along the way. In the end it worked (Martin L\"uscher remarked to me with something close to awe that the Kerman--Klein result agreed with the other two). But my bookkeeping did not generate, by itself, the graph counts.

About a month after I had submitted the paper, Jeffrey came into my office to show me that expanding a single exponential generated every graph to a given order, together with its count. It was beautiful. Using it, I checked my whole calculation in one afternoon. Neither he nor I ever published it, and I no longer remember the function in the exponent; what I remember is how little effort it took.

\section*{Data availability}

The notebook that produces every number in this note is archived on Zenodo~\cite{JacobsGWRimCode2026}.

\bibliography{gw_rim}

\end{document}